\documentclass[%
reprint,
superscriptaddress,
nofootinbib,
 amsmath,amssymb,
 aps,
 prc,
]{revtex4-1}

\usepackage[colorlinks,citecolor=blue]{hyperref}
\usepackage{graphicx}
\usepackage{dcolumn} 
\usepackage{bm}      
\usepackage{xcolor}
\usepackage{dcolumn}
\newcolumntype{d}{D{.}{.}{-1}}
\usepackage[colorinlistoftodos]{todonotes}

\newcommand{\fm}{\mathrm{\, fm}}
\newcommand{\MeV}{\mathrm{\, MeV}}

\renewcommand{\vec}[1]{\mbox{\boldmath $#1$}}

\begin{document}

\title{A Survey of Nuclear Pasta in the Intermediate Density Regime\\
I: Shapes and Energies}

\author{B. Schuetrumpf}
\affiliation{GSI Helmholzzentrum f\"ur Schwerionenforschung, Planckstra\ss{}e 1, 64291 Darmstadt, Germany}
\author{G. Mart\'{i}nez-Pinedo}
\affiliation{GSI Helmholzzentrum f\"ur Schwerionenforschung,
  Planckstra\ss{}e 1, 64291 Darmstadt, Germany} 
\affiliation{Institut f{\"u}r Kernphysik
  (Theoriezentrum), Technische Universit{\"a}t Darmstadt,
  Schlossgartenstra{\ss}e 2, 64289 Darmstadt, Germany}
\author{Md Afibuzzaman}
\affiliation{Computer Science and Engineering, Michigan State University
  East Lansing, Michigan 48824, USA}
\author{H. M. Aktulga}
\affiliation{Computer Science and Engineering, Michigan State University
  East Lansing, Michigan 48824, USA}

\date{\today}

\begin{abstract}

\begin{description}
\item[Background] Nuclear pasta, emerging due to the
  competition between the long-range Coulomb force and the short-range
  strong force, is believed to be present in astrophysical scenarios,
  such as neutron stars and core-collapse supernovae. Its structure
  can have a high impact e.g. on neutrino transport or the tidal
  deformability of neutron stars.

\item[Purpose] We study several possible pasta configurations, all of
  them minimal surface configurations, which are expected to appear in
  the mid-density regime of nuclear pasta, i.e. around 40\% of the
  nuclear saturation density. In particular we are interested in the 
  energy spectrum for different pasta configurations considered. 

\item[Method] Employing the density functional theory (DFT) approach, we calculate the binding energy of the different configurations for three values of the proton content $X_P=1/10,\,1/3$ and $1/2$, by optimizing their periodic length. We study finite temperature effects and the impact of electron screening.

\item[Results] Nuclear pasta lowers the energy significantly compared
  to uniform matter, especially for $X_P\geq1/3$. However, the
  different configurations have very similar binding energies. For
  large proton content, $X_P \gtrsim 1/3$, the pasta configurations
  are very stable, for lower proton content temperatures of a few MeV
  are enough for the transition to uniform matter. Electron screening
  has a small influence on the binding energy of nuclear pasta, but
  increases its periodic length.

\item[Conclusion] Nuclear pasta in the mid-density regime lowers the
  energy of the matter for all proton fractions under study. It can
  survive even large temperatures of several MeV. Since various
  configurations have very similar energy, it is to expect that many
  configurations can coexist simultaneously already at small
  temperatures.

\end{description}
\end{abstract}
\maketitle
\section{Introduction}
\begin{figure*}[htb]
    \centering
    \includegraphics[width=0.7\textwidth]{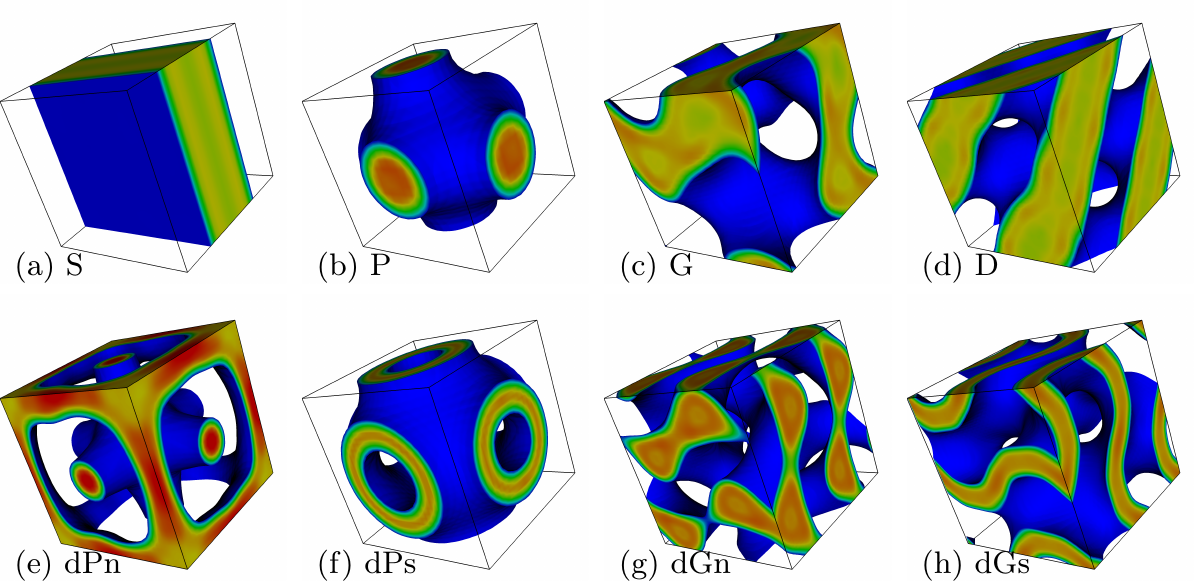}
    \includegraphics[width=0.1\textwidth, height=6cm]{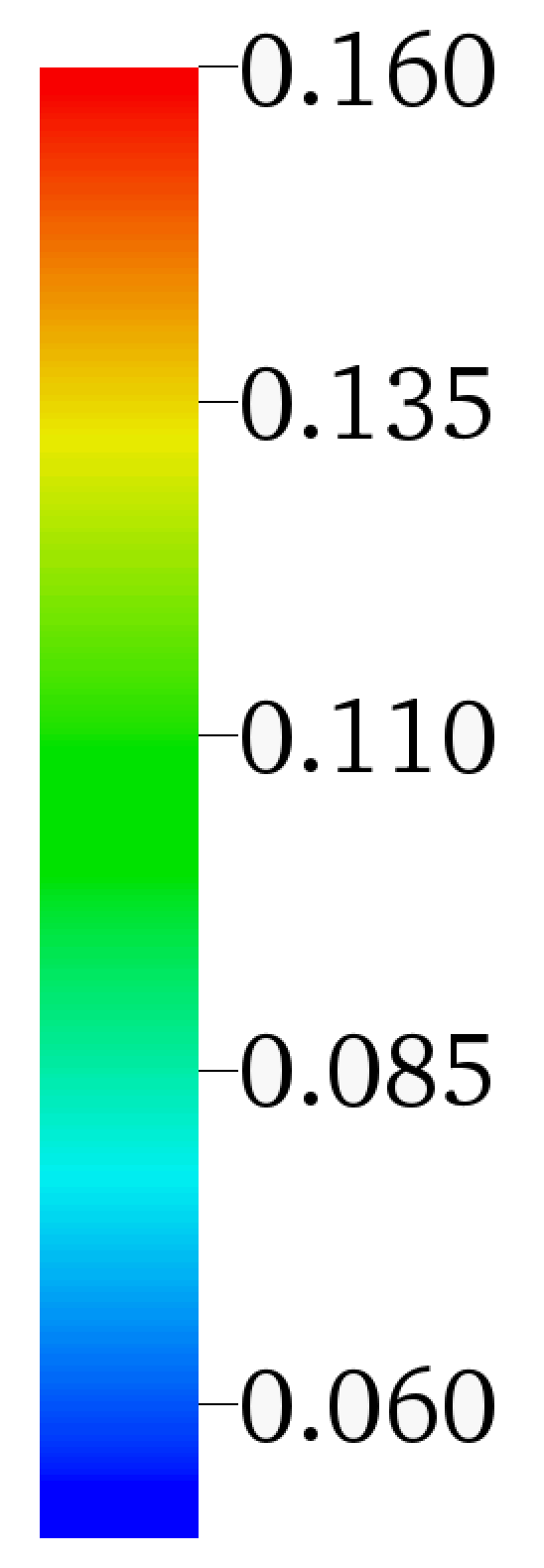}    
    \caption{One cubic elementary cell for all studied pasta configurations with $X_P=0.5$ and a mean density of $\rho=0.05\fm^{-3}$. Slab (a), P-surface (b), Gyroid (c), Diamond (d), network-like double P-surface (e), surface-like double P-surface (f), network-like double Gyroid (g), surface-like double Gyroid (h). The color scale is in units of $\fm^{-3}$.}
    \label{fig:Minimal_Surfaces}
\end{figure*}

The recent observation of gravitational waves from the neutron star
merger event GW170817~\cite{Abbott2017} and its electromagnetic
transient (AT~2017gfo)~\cite{Abbott.Others:2017}, which identifies
neutron star mergers as a site for the r process, has opened new
avenues to study the physics of matter in neutron
stars~\cite{Tews.Margueron.Reddy:2019,Gandolfi.Lippuner.ea:2019}. Gravitational
wave observations have been used to determine the tidal polarizability
(or deformability) of neutron stars and hence put limits on the
stellar radii and the underlying equation of
state~\cite{De.Finstad.ea:2018,*De.Finstad.ea:2018err,Abbott.Abbott.ea:2018}. In
addition to stellar compactness, the tidal polarizability is sensitive
to the second tidal Love number that has been shown to depend on the
inner crust of the neutron star~\cite{Piekarewicz.Fattoyev:2019}. 

The properties of the inner crust of neutron stars are affected by the
presence of nuclear pasta matter. Pasta matter is named for its
resemblance to Italian pasta, e.g. spaghetti and
lasagna~\cite{Ravenhall,Hashimoto} and is formed because of the
competition between the long-range Coulomb force and the short range
nuclear force (Coulomb frustration). It can appear at densities
between about 10\% and 90\% percent of the nuclear saturation density
and low enough temperatures. It is not possible to create an
environment for nuclear pasta in the laboratory. However, first
effects of Coulomb frustration can be observed in superheavy nuclei
\cite{Davies1972,Wong1973a,Dietrich1998,Decharge1999,Schuetrumpf2017a,Staszczak2017}.

A second site for nuclear pasta matter are core-collapse
supernovae. While in neutron stars the proton content of the nuclear
matter in the inner crust is expected to be about $X_P=1/10$ and
temperatures are low, the proton fraction in supernovae can be much
higher and temperatures can reach up to about 40~MeV. Nuclear pasta matter can have a strong influence on the neutrino
transport
\cite{Horowitz2004,Horowitz20042,Sonoda2007,Gry10,Roggero.Margueron.ea:2018}. Due
to its location in the neutron star, nuclear pasta can leave an
imprint on the neutrino spectrum of neutron stars. Furthermore,
elastic properties of nuclear pasta are different from uniform matter
or from spherical nuclei \cite{Caplan2018}.


Nuclear pasta has been studied with various approaches. Beside
classical theories, such as the liquid drop model and (quantum)
molecular dynamics calculations \cite{Dorso2012,Caplan2018} with which
it is possible to include vast numbers of nucleons and to simulate
very large systems, quantum theories, \emph{e.g.} the Thomas-Fermi
theory \cite{Williams1985,Oka13a,Pais15} and density functional theory
(DFT) \cite{Bonche,Mag02,Goe07a,NewtonStone,Pais12,
  Schuetrumpf2013,Schuetrumpf2014,Schuetrumpf2015,Schuetrumpf2015a,Fattoyev2017},
have also been employed. In these studies, it has been discovered that
not only the basic pasta structures such as the rod (spaghetti) or the
slab (lasagna) and their reversed (bubble) configurations
can be realized, but also much more complicated configurations. Among
them is the parking ramp configuration \cite{Berry2016}, similar to a
slab configuration but with defects, the Gyroid and the Diamond
\cite{Schuetrumpf2015,Nakazato2009,Nakazato2011}, and the P-surface
configurations \cite{Schuetrumpf2013} (see
Fig.~\ref{fig:Minimal_Surfaces}). The latter three are triply periodic
minimal surfaces (TPMS) and connect the research of nuclear pasta to
different fields, \emph{e.g.}, solid biological systems
\cite{MichielsenStavenga:2008,SchroederTurk2011}, di-block copolymers
\cite{Hajduk1994} and lipid-water systems \cite{Larsson:1989}.

In this work, we aim to give a comprehensive picture on pasta
configurations which are expected to form at the intermediate density
regime, $\rho\lesssim\rho_0/2$, including the TMPS, with the DFT
approach. In Sec.~\ref{sec:method}, we introduce our DFT
implementation, as well as TPMS and the method to extract observables
from the calculations. In Sec.~\ref{sec:GS}, we study the
configurations at zero temperature, in Sec.~\ref{sec:temp} we
introduce finite temperature and in Sec.~\ref{sec:screening} we
estimate the impact of electron screening. Finally, we give details of
the computational aspects in Sec~\ref{sec:scaling}.


\section{Method}\label{sec:method}
\subsection{The DFT approach}
In this work, the tool of choice to examine pasta configurations is
the DFT approach \cite{bender2003self}. We consider DFT at the
Hartree-Fock (HF) level, \emph{i.e.}, without pairing correlations
which would be accounted for by HF+BCS or full Hartree-Fock-Bogolyubov
(HFB) calculations. However they are not computationally feasible
for the large systems considered here. The DFT method can predict many
features, \emph{e.g.}, nuclear masses, radii or deformations of nuclei
all across the nuclear chart, and is therefore a valid tool also for
nuclear pasta.

We choose a standard Skyrme type parametrization of the DFT
functional. In particular, we choose the TOV-min parametrization
\cite{Erler2013} and use it throughout this work. This interaction is
not only fitted to stable nuclei and infinite matter properties, but
also to reproduce the mass-radius relation for neutron stars which
makes it very relevant for this work.

As engine to solve the DFT problem, we use the code Sky3D
\cite{Mar15a,Schuetrumpf2018}. It operates on a 3D equidistant grid in
a rectangular computational box. The derivatives are performed
utilizing the Fast Fourier Transform (FFT) technique which makes it easy
to implement periodic boundary conditions (PBC) for our pasta
calculations.  We assume periodic systems to overcome the mismatch in
size between the quantum system feasible to simulate on a
supercomputer involving a few thousand nucleons \cite{Afibuzzaman2018}
and the macrocopic length scale of a few hundred meters of the inner
crust of a neutron star. However, it has been shown that strict PBC
for the wave functions lead to spurious finite-volume
effects~\cite{Schuetrumpf2015a}. To reduce those errors, we employ the
twist-averaged boundary conditions (TABC)~\cite{Schuetrumpf2015a}. We
use TABC for zero-temperature calculations with a cubic box length of
$L\lesssim 24\fm$. For finite-temperature and large box lengths, we
have checked that PBC are sufficient due to the large number of
single-particle momentum states that are (partly) occupied.

\subsubsection{Finite temperature}

For finite-temperature calculations, we allow the states to be partly occupied. The density can be obtained by 
\begin{equation}
    \rho_{q}(\vec{r})=\sum_{\alpha}f_{\alpha,q}|\psi_{\alpha,q}(\vec{r})|^2
\end{equation}
where $q$ denotes the isospin (neutron or proton) and $\alpha$ the HF
single-particle state, $f_{\alpha,q}$ is the Fermi distribution for
neutrons and protons separately. Chemical potentials are determined
from the particle number constraint. The sum over HF single-particle
states is such that for the highest state in energy
$f_{\alpha,q}<0.01$. Instead of minimizing the internal energy of the
system, for finite-temperature calculations the HF technique minimizes
the free energy $F=E-TS$. The entropy of the system can be obtained
from
\begin{equation}
    S=-\sum_{\alpha,q} \left[ f_{\alpha,q}\ln(f_{\alpha,q})+(1-f_{\alpha,q})\ln(1-f_{\alpha,q})\right]\quad .
\end{equation}

\subsubsection{Electron screening}
In Sec.~\ref{sec:screening}, we will estimate the effect of a non-uniform electron distribution on nuclear pasta. Electrons will be attracted by the positive charge distribution coming from the protons and effectively screen the Coulomb potential from the protons.

\begin{figure}[!t]
    \centering
    \includegraphics[width=\columnwidth]{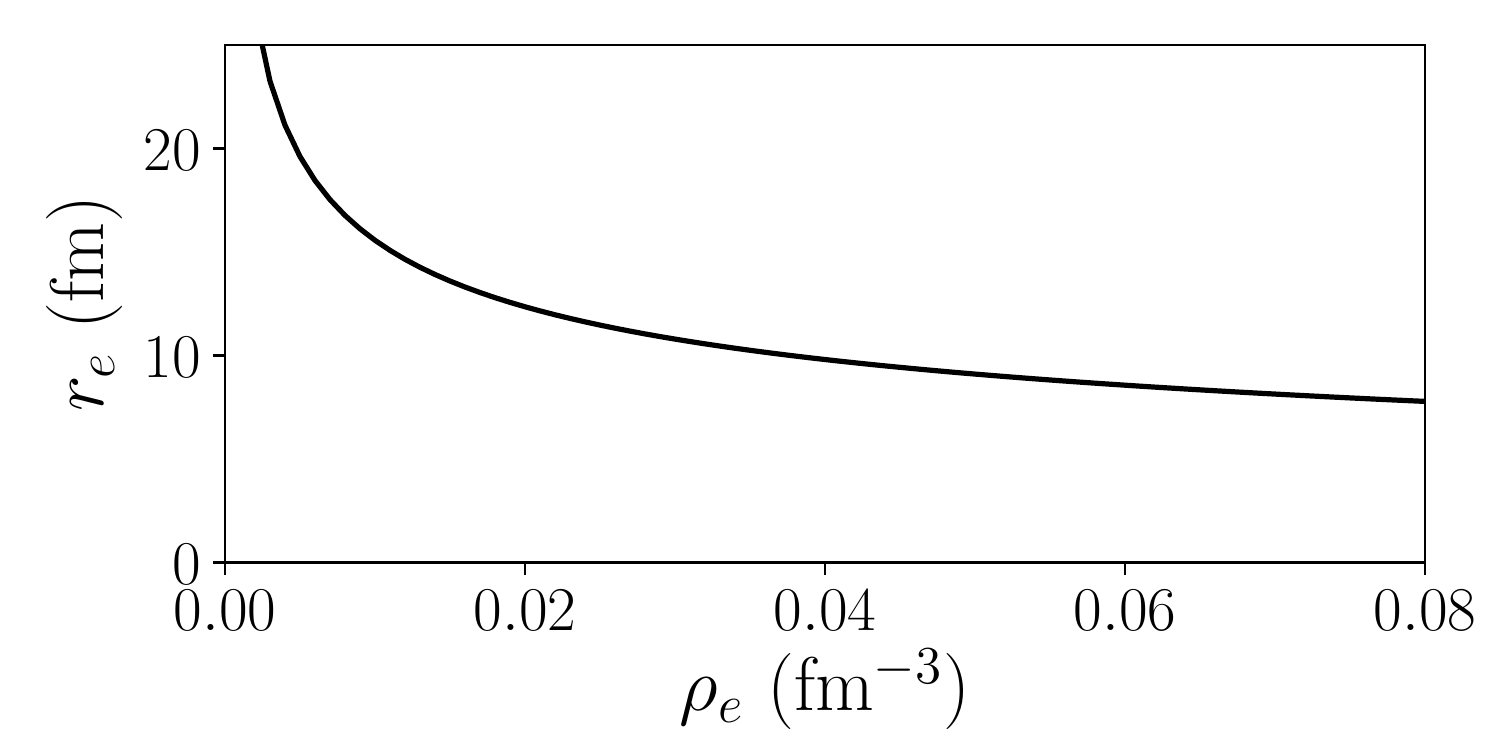}
    \caption{Electron screening radius as a function of mean electron density.}
    \label{fig:screening_length}
\end{figure}

To account for electron screening we use the Thomas-Fermi
approximation that leads to the Poisson equation~\cite{Haensel}
\begin{equation}
    \left(\vec{\nabla}^2+k^2_{\rm TF}\right)\phi(\vec{r})=-4\pi\rho_p(\vec{r})\quad,
\end{equation}
taking the charge distribution as the proton distribution $\rho_p$. It
can easily be solved in Fourier space 
\begin{equation}
    \tilde{\phi}(\vec{k})=\frac{-4\pi\tilde{\rho}_p(\vec{k})}{(-\vec{k}^2+k^2_{\rm
        TF})}\quad. 
\end{equation}
The Thomas-Fermi wave number is
\begin{equation}
 k_{\rm TF}=r_e^{-1}=\left(4\pi e^2 \frac{\partial \rho_e}{\partial \mu_e}\right)^{1/2}
\end{equation}
and its inverse, $r_e$, the electron screening length. For a strongly
degenerate electron gas, which we consider here in this work, the
Thomas-Fermi wave number can be expressed as \cite{Haensel} 
\begin{equation}
    k_{\text{TF}}=2\sqrt{\frac{\alpha_f\sqrt{1+x_r^2}}{\pi x_r}}k_\mathrm{F}
\end{equation}
where $x_r=\hbar k_\mathrm{F}/(m_ec)$, $\alpha_f$ is the fine
structure constant and $k_\mathrm{F}=(3\pi^2 \rho_e)^{1/3}$ is the
Fermi momentum of the electrons. The mean electron density, $\rho_e$,
due to charge neutrality is the same as the mean proton density. The
resulting electron screening length as a function of the mean electron
density is shown in Fig.~\ref{fig:screening_length}.

\subsection{Minimal Surfaces for Nuclear Pasta}
The goal of this work is to determine the binding energy of nuclear
pasta in the intermediate density regime where the configurations
shown in Fig.~\ref{fig:Minimal_Surfaces} are expected to appear. To
that end, we have to find the optimal pasta configuration for a given
mean density, proton fraction and temperature. Generally, all possible
configurations have to be assumed and from those the one with minimal
energy is considered as the ground state. In this work, we only
consider the periodic configurations shown in
Fig~~\ref{fig:Minimal_Surfaces}. The simplest among them is the slab
(S) configuration which has already been studied from the very
beginning of pasta matter research. The others we consider are the
Schwarz Primitive (P-)surface, Diamond (D-)surface, and Gyroid
(G-)surface. Their nodal approximations are
\begin{subequations}
\begin{align}
     \phi_S &= \cos X  \\
     \phi_P &= \cos X + \cos Y + \cos Z \\
     \phi_G &= \cos X \sin Y + \cos Y \sin Z + \cos Z \sin X\\
     \phi_D &= \cos X \cos Y \cos Z + \cos X \sin Y \sin Z \nonumber\\
            &+ \sin X \cos Y \cos Z + \sin X \sin Y \cos Z\quad ,
\end{align}
\end{subequations}
where $X=2\pi x/L$ and likewise for the other directions.

In first order, the minimal surfaces can be parametrized with the
nodal approximations as $\phi_i=0$ with $i\in\{S,P,G,D\}$. The surface
divides the space into two completely separated half spaces with equal
volume and equal topology. Only the Gyroid divides the space into two
half spaces with opposite chirality. When the half spaces are not
divided equally, we can approximate the surface as $\phi_i=t$
for small values of $t$. If one half space or domain is filled with
nuclear matter and the other is empty or only filled with a neutron
background gas, we call those configurations ``single''
configurations. ``Double'' configurations are enclosed by two surfaces
$\phi_i=\pm t$. For double configurations there are two possibilities:
Either the surface-like domain $|\phi_i|<t$ or the network-like domain
$|\phi_i|>t$ can be filled with nuclear matter. Henceforth, they are
labeled as e.g. dGs for double Gyroid surface-like and similar for the
other double configurations. We do not consider the double Diamond
configurations, because their preferred periodic lengths are too large
and computationally not feasible.

\begin{table}[htb]
    \centering
    \begin{ruledtabular}
    \begin{tabular}{lcccc}
    &S&P&G&D  \\\hline
    $\chi$&0&-2&-4&-8\\
    $A_1$&2.0&2.15652&2.65624&3.3715\\
    \end{tabular}
    \end{ruledtabular}
    \caption{Euler characteristic $\chi$ per cubic unit cell and
      surface area $A_1$ for an assumed box length of 1 for single
      minimal surface configurations.}
    \label{tab:Min_surf}
\end{table}

The TPMS (P, G, and D) are related to each other by the Bonnet
transformation, which transforms the shapes smoothly into each other
\cite{HydeLanguageOfShape:1997}. Our three candidates are the only
physically relevant surfaces, since all others are
self-intersecting. A difference between the surfaces is that they have
different surface areas for the same periodic length 
\begin{equation}
    A_i=A_{1,i}\times L^2,\quad i\in\{S,P,G,D\}.\label{eq:surface}
\end{equation}
The configurations also have different Euler characteristics for one
cubic unit cell, which is often used to discriminate pasta
configurations. It is defined as  
\begin{equation}
    \chi=\#(\text{connected components})-\#(\text{holes})+\#(\text{cavities})
\end{equation}
The values for $A_{1,i}$ and $\chi$ for a cubic elementary cell are given
in Table~\ref{tab:Min_surf} \cite{Schroeder2003}.

\subsection{Minimizing the pasta binding energy}

For each configuration, we perform DFT calculations for a fixed
average density and proton fraction in a given computational box which
is equal to the periodic length of the configuration. By varying the
periodic length, we determine the energy of each configuration as a
function of periodic length. The binding energy corresponds to the
energy minimum that in general is obtained at different optimal
periodic lengths for each configuration.

In order to computationally define a certain configuration, we fix the
mean-field for the first 200 iterations of the DFT calculations. For
the single configurations we take a guiding potential of the form of
the nodal approximations, i.e.
\begin{equation}
    U_{{\rm single,\,}it<200}=\phi_0\cdot \phi_i, \quad i\in\{S,P,G,D\},
\end{equation}
where $\phi_0$ is a constant optimized to speed up
convergence. Consequently, the matter will arrange itself in the
domain, where $\phi_i(\vec{r})<0$.

For the double configurations, we take the guiding potential of the form
\begin{equation}
    U_{{\rm double,\,}it<200}=\pm\phi_0\cdot |\phi_i|, \quad i\in\{P,G\},
\end{equation}
where the positive sign leads to the surface-like configuration and the negative sign leads to the network-like configuration. 

After 200 iterations, we perform standard DFT calculations at the HF
level with the self-consistent potential and find the local
minimum. Eventually, calculations yield the desired configuration. If
the actual box length is far off the optimal length, the configuration
can be unstable and will then be transformed into some other
configuration. Those cases are not considered.

\begin{figure}[htb]
    \centering
    \includegraphics[width=0.9\columnwidth]{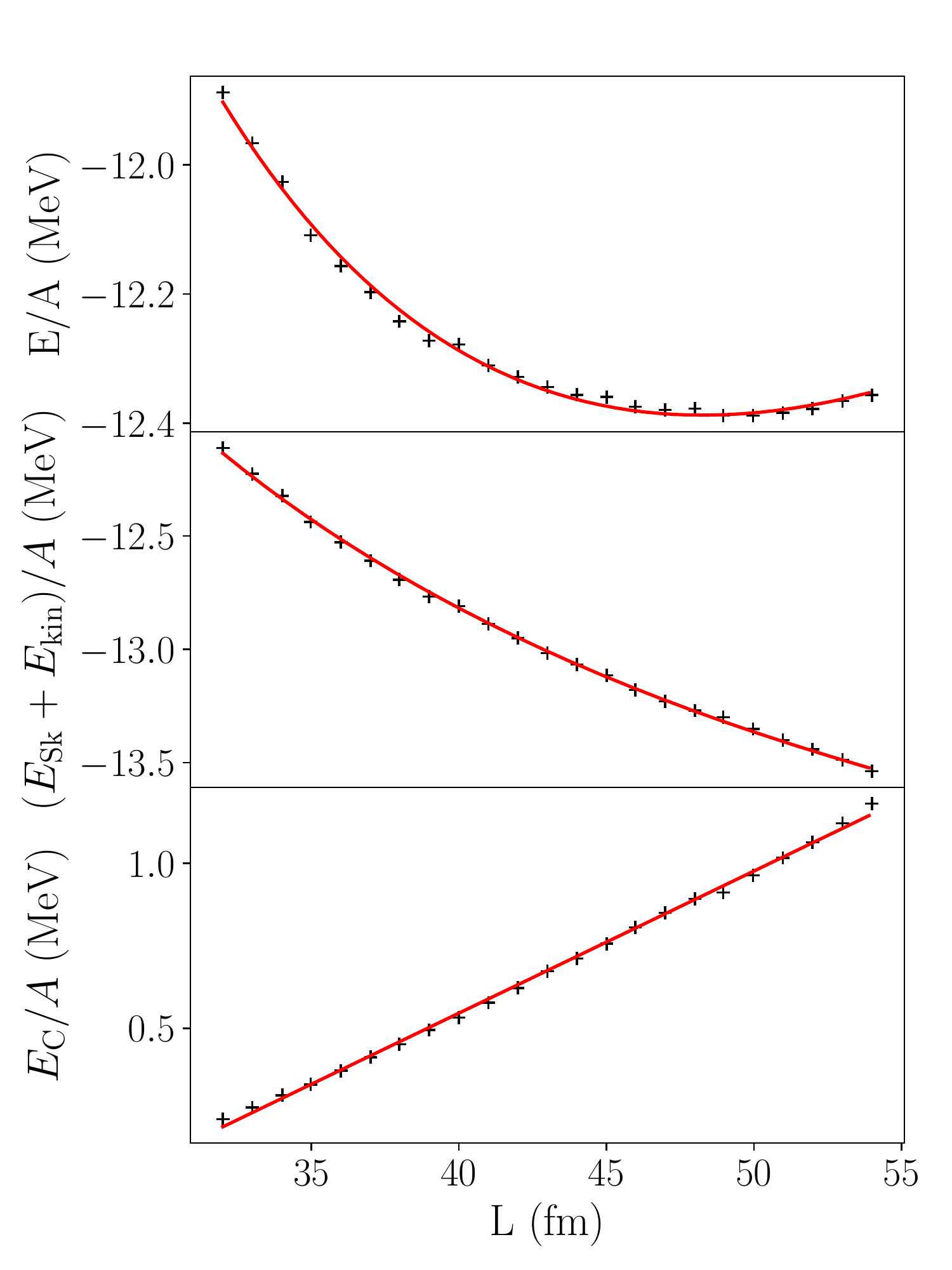}
    \caption{Total energy per nucleon, Skyrme energy and kinetic energy per nucleon, and Coulomb energy per nucleon as a function of the cubic periodic length $L$ for the dGs configuration with $X_P=0.5$ and $\rho=0.05\fm^{-3}$. The black + signs correspond to the HF calculations, the red line displays the fit to the calculations.}
    \label{fig:opt_E}
\end{figure}

In order to determine the binding energy and optimal periodic length
we fit the energy as a function of periodic length for the discrete
computed values using a functional form based on the following
assumptions.  We divide the total energy into two parts: Skyrme
(potential) energy plus kinetic energy and Coulomb energy containing
both the direct and exchange contributions. As a function of the
periodic length, Coulomb energy per particle behaves approximately
linearly in the region around the optimal periodic length of the
configuration. We base this assumption on empirical observation. In a
liquid drop picture, the energy per particle contains two leading
terms: a volume term that is constant and a surface term that behaves
like $\sim L^{-1}$. In summary, we fit the parts with

\begin{subequations}
\begin{alignat}{4}
     (E/A)_\mathrm{fit} &=   aL   &&+  && bL^{-1}   &&+   c  \\
     ((E_\mathrm{Sk}+E_\mathrm{kin})/A)_\mathrm{fit} &=   && &&   bL^{-1}   &&+   d  \\
     (E_C/A)_\mathrm{fit} &= aL && && &&+ e,
\end{alignat}
\end{subequations}
where $c=d+e$. An example for the dGs configuration is shown in
Fig.~\ref{fig:opt_E}. We determine the parameters $a, b$, and $c$ by
directly fitting the total energy. Once they are known, the optimal
periodic length and minimum energy are determined from the
expressions: $L_\mathrm{opt} =\sqrt{b/a}$, $(E/A)_\mathrm{opt}=2\sqrt{ab}+c$. 

For the production runs, we perform only 5 to 7 calculations around
the optimal box length in steps of $\Delta L=2\fm$ for the shapes with
smaller periodic length and $\Delta L=4\fm$ for the shapes with larger
periodic length. We have estimated that this procedure gives lengths
with an error of about $1\fm$ and energies with an error below $0.01\MeV$.


\section{Zero Temperature Pasta Matter}\label{sec:GS}
\begin{figure}[htb]
    \centering
    \includegraphics[width=\columnwidth]{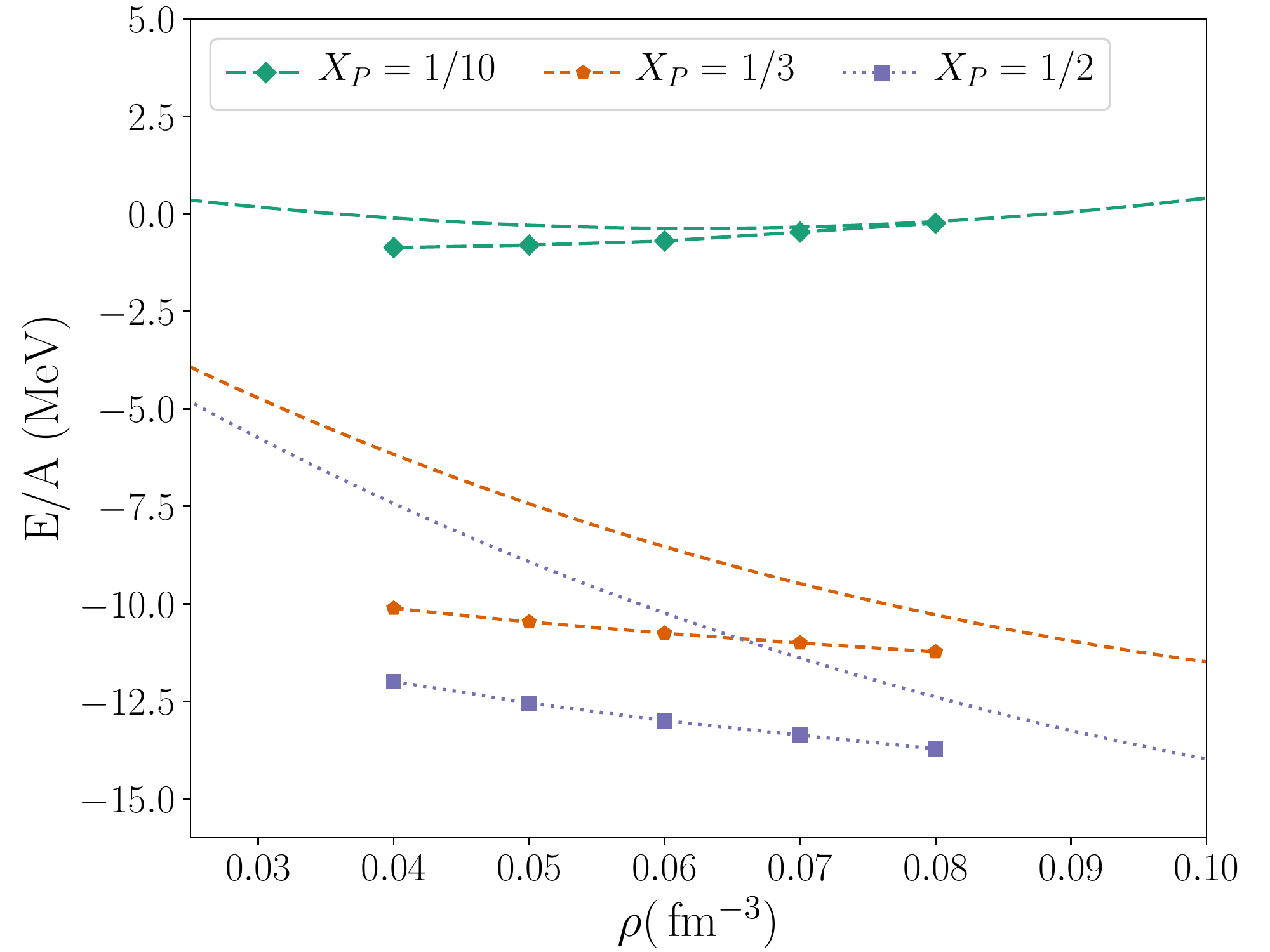}
    \caption{Comparison of total energies per nucleon of uniform matter (dashed lines) and slab (S) configuration (dashed lines with symbols) in the intermediate density regime for different proton fractions.}
    \label{fig:Slab_Uni_E_zero}
\end{figure}

We consider the pasta configurations shown in
Fig.~\ref{fig:Minimal_Surfaces} in the density region between
$0.04\fm^{-3}$ and $0.08\fm^{-3}$ in steps of $0.01\fm^{-3}$. They are
calculated for three different values for the proton content:
$X_P\in \{1/10, 1/3, 1/2\}$. Depending on the computational box 
and mean density, the calculations involve between a few hundred 
up to several thousand nucleons. The computational cost is briefly discussed in Sec.\ref{sec:scaling}. 

Fig.~\ref{fig:Slab_Uni_E_zero} compares the ground state energies of
the slab (S) configuration with optimal periodic length to uniform
matter. For $X_P=1/2$ and $1/3$ at low densities the S pasta
configuration is lower in energy for as much as $5\MeV$ per
particle. At the highest densities studied ($0.08\fm^{-3}$) the
difference is about $1\MeV$. For $X_P=1/10$ the energy for low
densities is lowered by about $1\MeV$ and approaches the uniform
matter total energy per particle for $0.08\fm^{-3}$.

\begin{figure}[htb]
    \centering
    \includegraphics[width=\columnwidth]{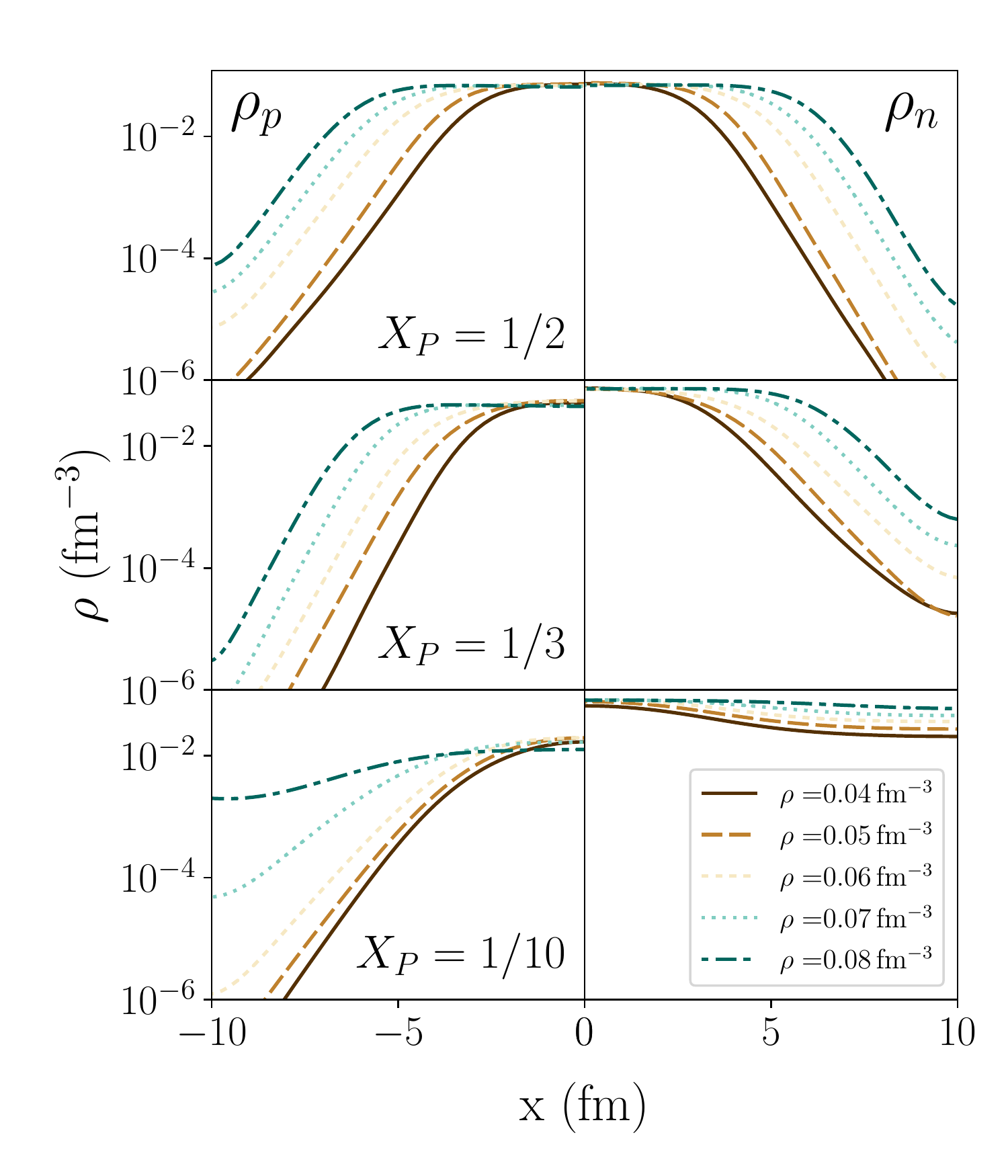}
    \caption{Density profiles for the slab configuration in the
      non-uniform direction. Proton densities are on the left, neutron
      densities are on the right panels. Symmetric nuclear matter is
      in the upper panel, $X_P=1/3$ in the mid panels and $X_P=1/10$
      in the lower panels.} 
    \label{fig:rho_slab}
\end{figure}

Slab density profiles for all studied mean densities and proton
fractions are shown in Fig.~\ref{fig:rho_slab}. These calculations
were done for a fixed periodic length of $L=20\fm$. For symmetric
nuclear matter the profiles for neutrons and protons look similar. The
protons have a slightly larger background density in the low density
region (or void) between the slabs than the neutrons due to Coulomb
repulsion. For larger mean densities the slabs extend more and the
void region becomes smaller. Therefore the minimum density in the void
region becomes larger, up to $10^{-4}\fm^{-3}$ for protons for a mean
density of $0.08\fm^{-3}$.

For the other proton fractions considered, the maximum density inside
the slab is much larger for neutrons than for protons similar to what
is found in finite nuclei~\cite{Schuetrumpf2017a}. While the proton
densities become very low in the void region ($10^{-6}\fm^{-3}$ and
below), some neutrons are evaporated and an appreciable neutron
background is visible for $X_P=1/3$. For $X_P=1/10$ the transition to
uniform matter is visible with increasing density, explaining the
convergence of the energies to the uniform matter limit
in~Fig.~\ref{fig:Slab_Uni_E_zero}. For the lowest densities the slab
shape is still clearly visible in the proton densities.

\begin{figure}[htb]
    \centering
    \includegraphics[width=\columnwidth]{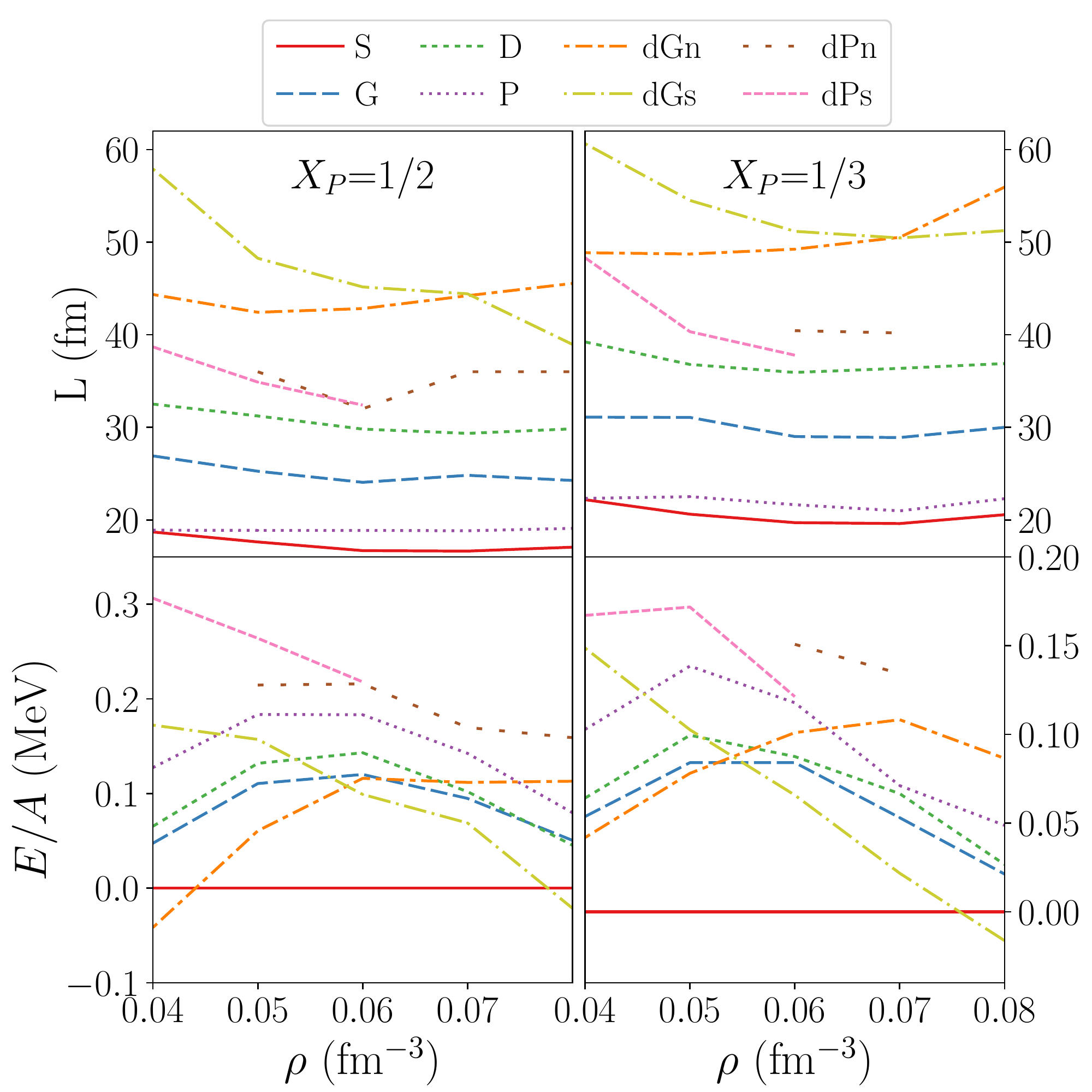}
    \caption{Optimal periodic lengths (top panels) and binding
      energies per particle shifted with respect to the slab shape
      (bottom panels) for minimal surface configurations. Left panels:
      $X_P=1/2$, right panels: $X_P=1/3$.}
    \label{fig:Pasta_E-L_zero}
\end{figure}

Fig.~\ref{fig:Pasta_E-L_zero} shows the optimal periodic lengths and
the corresponding binding energies with respect to the binding energy
of the slab configuration. It has to be mentioned that for both double
P topologies, stable configurations could only be obtained for a
limited mean density range. The surface-like structure was stable for
low densities and the network-like structure was stable for higher
energies, except for $X_P=1/3$ and $\rho=0.08\fm^{-3}$. The unstable
configurations (not shown in Fig.~\ref{fig:Pasta_E-L_zero}) were
transformed mostly to nuclei arranged in a bcc lattice at lower
densities and to nuclear bubbles at higher densities. For very large
box lengths, the dPn configuration formed extra
cavities at the knots, which can slightly be
seen already in Fig.~\ref{fig:Minimal_Surfaces}, panel (e), where
yellow lower density regions appear at the corners of the box. Those
shapes were not considered, as they are topologically different and
out of scope of this work.

We only show the results for $X_P=1/2$ and 1/3. For $X_P=1/10$, the
binding energies barely depends on the periodic length. For all the
configurations the binding energy is very close to the one of the
slab, especially at high mean densities. 
For $X_P\gtrsim 1/3$ none of the binding energies
differ more than $0.3\MeV/A$ from the one of the slab configuration,
most of them differ not more than $0.1\MeV/A$. The slab is, for most
of the density range, the configuration with the lowest energy. Only
for $\rho=0.4\fm^{-3}$ and $\rho=0.8\fm^{-3}$ the double gyroids, dGn or dGs,
have lower energy. However at those densities different shapes come
into play~\cite{Schuetrumpf2013}, such as the rod, anti-rod or waffle configurations, which
are not discussed in this work.

\begingroup
\squeezetable
\begin{table}[t]
  \centering
    \begin{ruledtabular}
  \renewcommand{\arraystretch}{1.3}
    \begin{tabular}{ldddd@{\hskip 0.5cm}dddd}
      $X_P$&\multicolumn{4}{c}{1/2}&\multicolumn{4}{c}{1/3} \\ \cline{2-5}\cline{6-9}
      shape&\multicolumn{1}{c}{S}&\multicolumn{1}{c}{P}&\multicolumn{1}{c}{G}&\multicolumn{1}{c}{D}&\multicolumn{1}{c}{S}&\multicolumn{1}{c}{P}&\multicolumn{1}{c}{G}&\multicolumn{1}{c}{D}\\\hline
      $\bar{L}$&17.33&18.91&25.06&30.54&20.53&21.94&30.01&37.04\\
      $A/V$ & 1.16&1.14&1.06&1.1&0.97&0.98&0.89&0.91\\
      $\chi/V$&0&2.96&2.54&2.81&0&1.89&1.48&1.57\\ 
    \end{tabular}
    \end{ruledtabular}
    \caption{Optimal box length, averaged over all densities $\bar{L}$ in $\fm$, ratio of surface to volume $A/V$ in $10^{-1}\fm^{-1}$ and Euler characteristic per volume in $10^{-4}\fm^{-3}$. The surface area $A$ is estimated with Eq.~(\ref{eq:surface}) using $\bar{L}$ and parameters in Table~\ref{tab:Min_surf}. Volume $V$ is the unit cell box volume $V=\bar{L}^3$.}
    \label{tab:Min_surf_ratio}
\end{table}
\endgroup

The optimal periodic lengths of the single shapes (S,P,G,D) are almost
constant in the range of mean densities under study. It is interesting
to note that the ratio of surface to volume is almost constant for all
four singe shapes (see Table~\ref{tab:Min_surf_ratio}). This results
also in an almost equal Euler characteristic per volume, except for
the slab shape, where the Euler characteristic is always zero. In
contrast, the optimal periodic lengths for surface-like double
structures decrease with increasing mean density and for the
network-like shapes we observe the opposite behaviour. The optimal
periodic lengths for double configurations are significantly larger
than those for the single configurations.

The single TPMS configurations show parabolic behavior w.r.t. the
slab configuration. The maximum is around $0.06\fm^{-3}$ for $X_P=1/2$
and slightly shifted to lower densities for $X_P=1/3$. The energy per
particle is decreasing for surface-like configurations with increasing
mean density and tendentiously increasing for network-like
configurations. While the double gyroid has lower energy than all of
the single TPMS configurations (the network-like structure for lower
energies and surface-like for higher energies), the double P
configurations are both higher in energy.


\section{Finite Temperature}\label{sec:temp}

In this section, we study the impact of temperature on pasta
shapes. As a representative configuration, we chose the slab
configuration because it is the most bound or ground state
configuration. Thus, the temperature at which the slab melts
represents the disappearance of the pasta phases. We restrict our
investigation to the lowest mean density considered in
section~\ref{sec:GS}, $\rho=0.04\fm^{-3}$. At this density, we find
the largest difference in binding energy between pasta phases and
uniform matter (see Fig.~\ref{fig:Slab_Uni_E_zero}) and hence the
largest pasta phase melting temperature.

\begin{figure}[htb]
    \centering
    \includegraphics[width=\columnwidth]{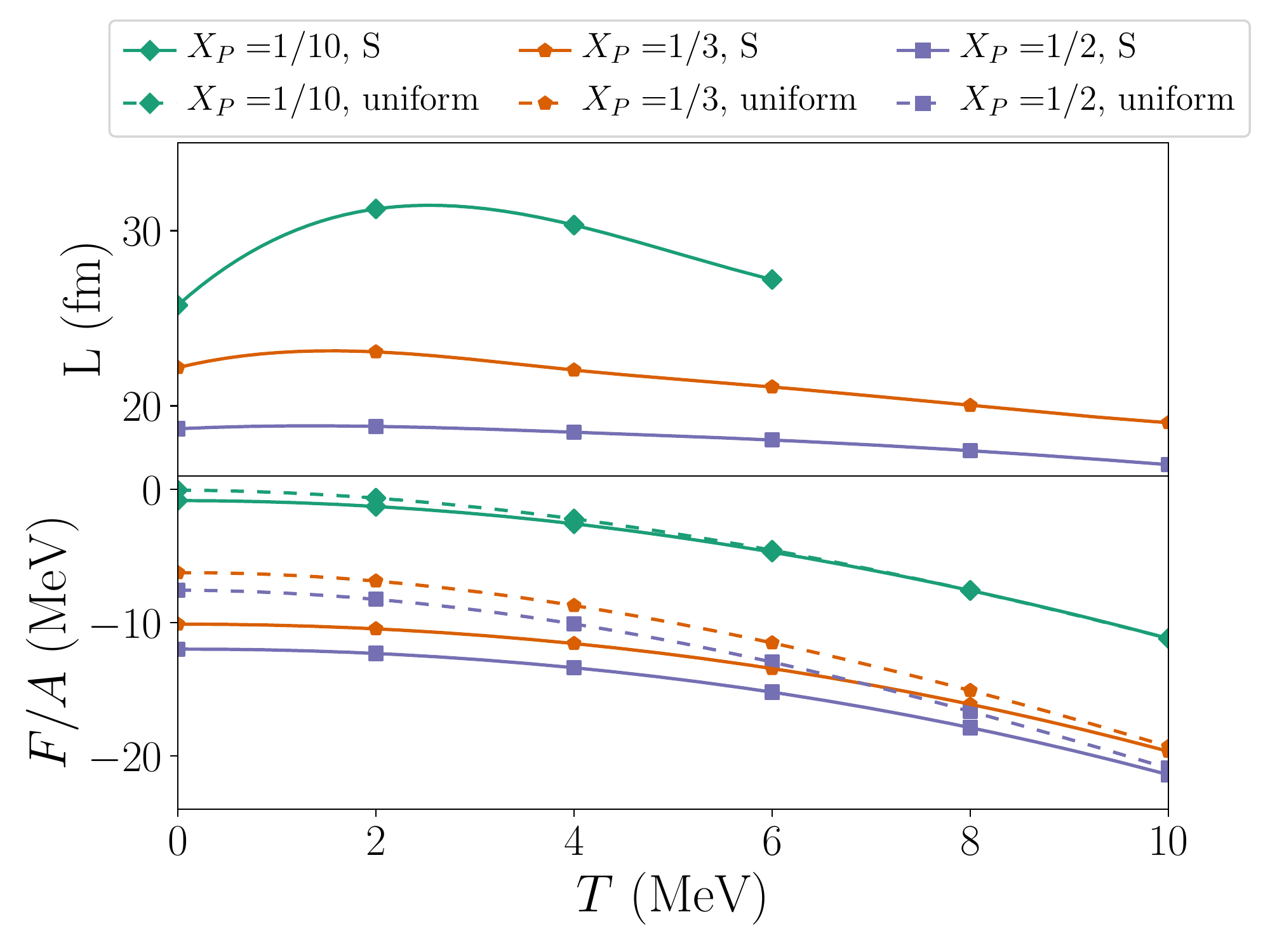}
    \caption{Optimal box length (top) and free energy of the slab at a
      mean density of $0.04\fm^{-3}$ with respect to temperature. The
      free energy is compare to the free energy of uniform matter
      (dashed lines). The connecting lines are simply cubic spline
      interpolations.} 
    \label{fig:temperature_E}
\end{figure}

Figure~\ref{fig:temperature_E} illustrates the optimal periodic
lengths and free energy per nucleon for temperatures up to
$10\MeV$. It is interesting to see that the preferred box length
increases for a temperature of $2\MeV$ and then decreases for higher
temperatures for all choices of $X_P$. For $X_P=1/10$ and
$T\geq 8\MeV$, the density distribution becomes very close to uniform
matter and therefore the optimal box length is not shown anymore.

While for zero temperature the internal or free energy for the slab and uniform matter differs by about $4.4\MeV$ for $X_P=1/2$ and $3.9\MeV$ for $X_P=1/3$, the difference for $X_P=1/10$ is only about $0.8\MeV$. At $T=10\MeV$, the difference shrinks to $0.5\MeV$ for $X_P=1/2$ and $0.35\MeV$ for $X_P=1/3$ and for $X_P=1/10$ the transition to uniform matter has already occurred between $T=6\MeV$ and $T=8\MeV$.

\begin{figure}[!t]
    \centering
    \includegraphics[width=\columnwidth]{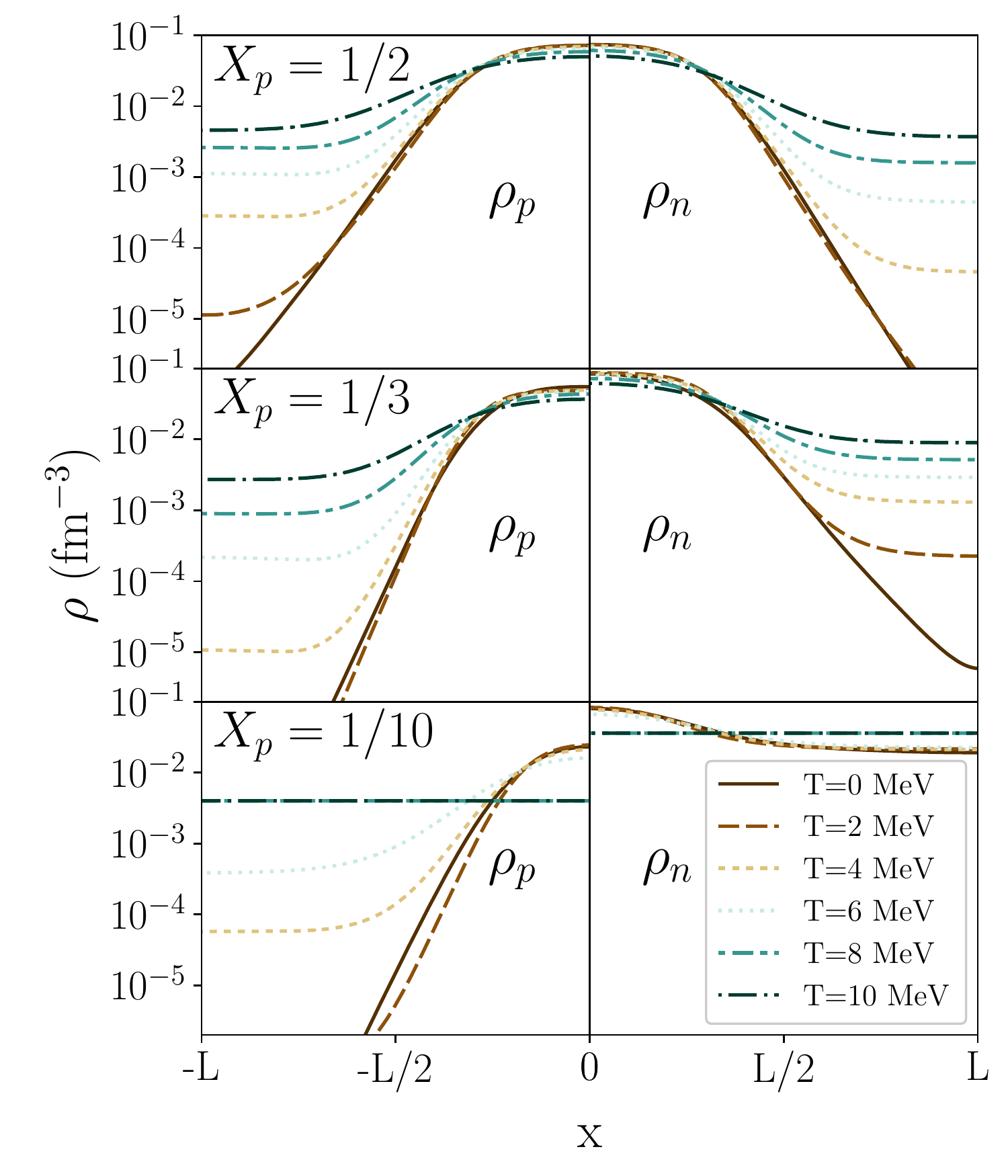}
    \caption{Density profiles for slabs at a mean density of $0.04\fm^{-3}$ with $X_P=1/2$ (top), $X_P=1/3$ (middle), $X_P=1/10$ (bottom) and varying temperature. Proton density is shown on the left, neutron density on the right.}
    \label{fig:slab_temperature}
\end{figure}

In Fig.~\ref{fig:slab_temperature}, the density profiles for the slabs for all investigated proton fractions and temperatures are shown. Note that the spatial direction is scaled according to the optimal box lengths to fit into the same plot. It is interesting to note that while increasing the temperature to $T=2\MeV$, the fraction of the surface area between the nuclear matter phase and the void or background phase becomes smaller. Going to higher temperatures, the transition is much more smooth and the surface area increases. At $T=10\MeV$, the densities of the $X_P=1/2$ and $1/3$ slabs are still clearly distinguishable from uniform matter. The transition takes place at still higher temperatures.


\section{Electron Screening}\label{sec:screening}
In the previous sections, we assumed that electrons form a uniform
background. However, electrons in this very dense environment can have
an influence and gather in the areas with high proton charge
density. Thus the electrons partly shield the Coulomb potential from
the protons and reduce the Coulomb energy. This phenomenon is called
electron screening.

\begin{figure}[tbh]
    \centering
    \includegraphics[width=\columnwidth]{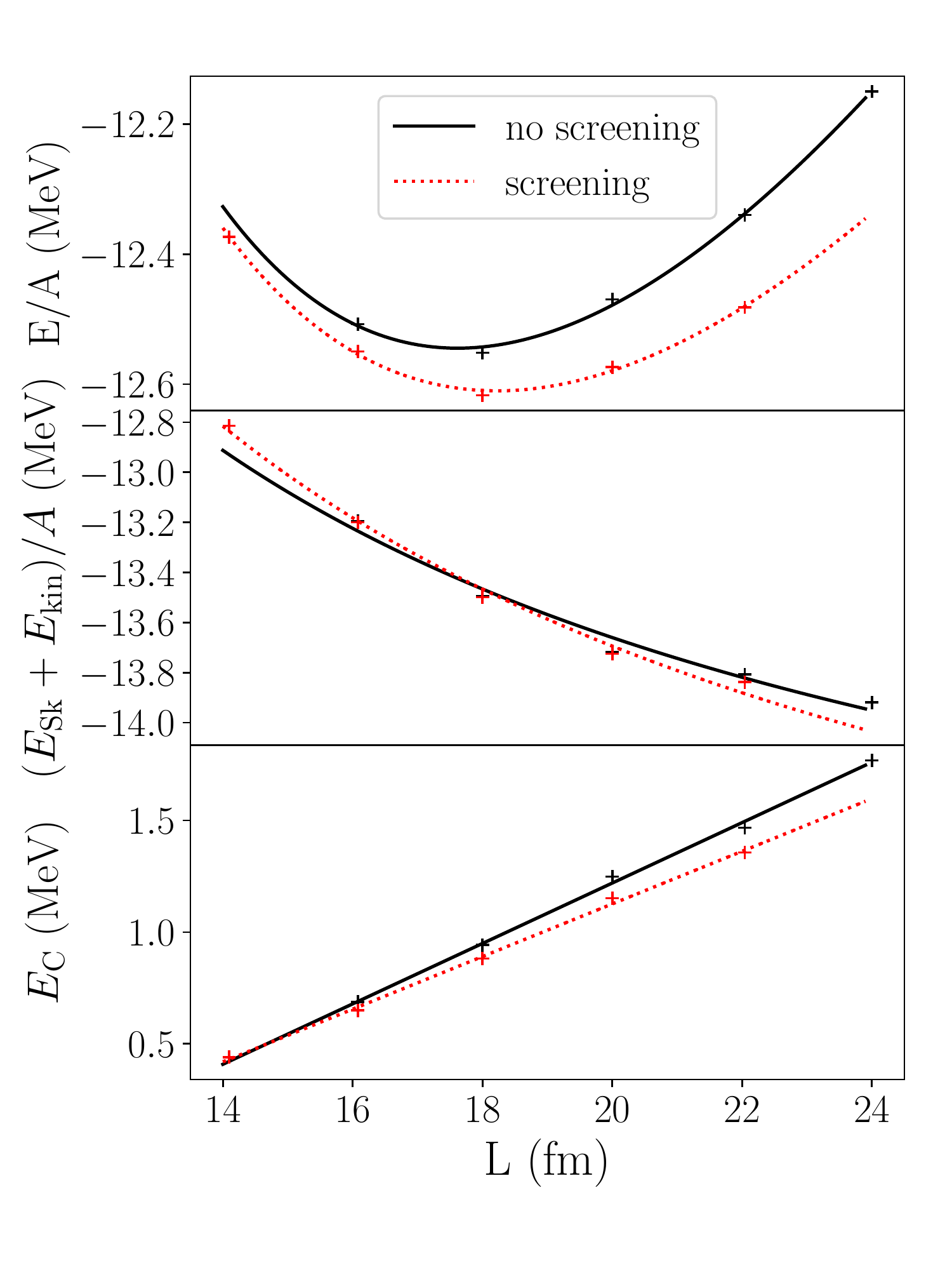}
    \caption{Total energy (top), Skyrme and kinetic energy (middle), and Coulomb energy (bottom) per nucleon for a slab configuration with $\rho=0.05\fm^{-3}$ and $X_P=1/2$ without electron screening (black, solid line) and with electron screening (red, dotted line).}
    \label{fig:E_screening}
\end{figure}

In Fig.~\ref{fig:E_screening}, the comparison of calculations with and
without screening is shown for $\rho=0.05\fm^{-3}$ and
$X_P=1/2$. While the Skyrme and kinetic energy part only varies
little, the Coulomb energy is systematically reduced after switching
the screening effect on. This causes the total energy to also be
reduced slightly. For the optimal periodic length, the total energy
changes from $-12.54\MeV$ to $-12.61\MeV$, which is about a 0.4\%
reduction. The optimal periodic length itself changes from $17.61\fm$ to
$18.22\fm$, which is a much larger relative change of 2.3\%.

For the most extreme case with $\rho=0.08\fm^{-3}$ and $X_P=1/2$, we
observe a comparable change in energy and a larger increase of the periodic
length of 4.8\%. We expect the impact to be much smaller for low
proton fractions because of the smaller charge density and thus a much
larger screening radius (see
Fig.~\ref{fig:screening_length}). Overall, we expect that the
influence for the binding energy is negligible and an increase of the
preferred periodic length of no more than 5\%.


\section{Computational Scaling}\label{sec:scaling}
Extensive calculations have been performed for this work, and, to the
best of our knowledge, they represent the largest nuclear DFT ground
state calculations ever performed. While most of the calculations were
executed on the LOEWE facility at the Goethe University Frankfurt,
calculations for the plots of the computational scaling shown below
were performed on Cori at NERSC (National Energy Research Scientific
Computing Center) in Berkeley, California. Cori - Phase I is a Cray
XC40 supercomputing platform that consists of 4766 compute nodes, each
with two 16-core Xeon E5-2698 v3 Haswell CPUs.  

\begin{figure}[!t]
    \centering
    \includegraphics[width=0.95\columnwidth]{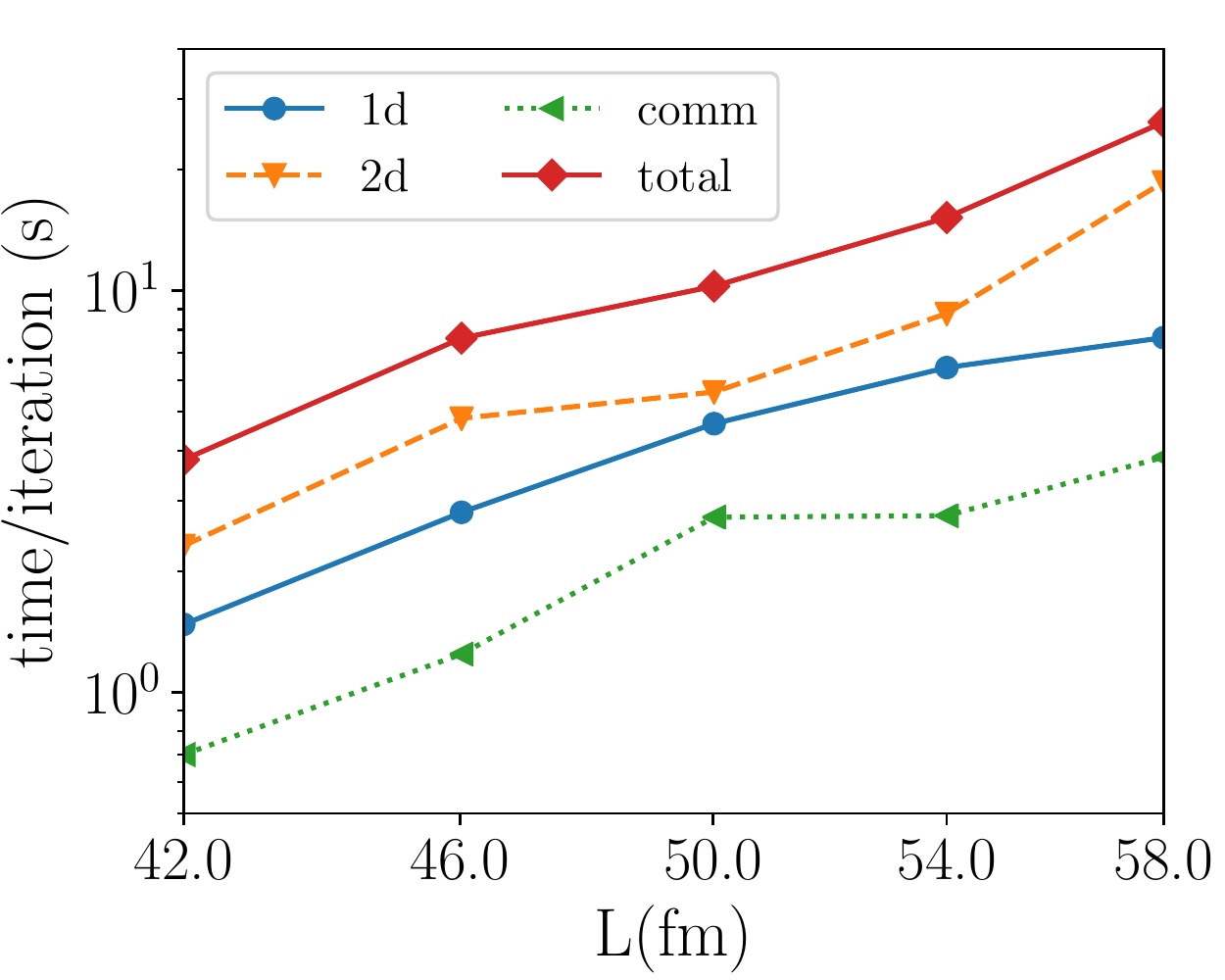}
    \caption{Total execution time per iteration using 512 cores for a
      system with $X_P=1/3$, $\rho=0.04$ with varying box lengths. The
      1d part stands for evolving the wave function with the damped
      gradient step and the 2d part stands for the orthonormalization
      and diagonalization of the Hamiltonian. ``Comm" marks the
      communication overhead. For details see \cite{Afibuzzaman2018}}.
    \label{fig:scaling_L}
\end{figure}

First we present the scaling results with a fixed core count of 512
for a system with $X_P=1/3$ and $\rho=0.04$ for different box sizes in
Fig.~\ref{fig:scaling_L}. We take a cubic box with 40 grid points in
each direction for $L=42\fm$, 44 for $L=46\fm$, 48 for $L=50\fm$ and
50 for the larger boxes. We would expect a linear scaling with total
grid points, if the number of wave functions was constant, because the
FFT part, scaling with $N\log(N)$, does not take a considerable amount
of time. The number of wave functions is determined by the density and
thus increases as $V=L^3$.

The total computational time per iteration is divided into three parts. 
``2d" labels the part where the data (wave functions, matrices, etc.) 
are distributed over the CPU cores in a block cyclic 2d way. In this 
distribution the wave functions are orthonormalized and diagonalized 
w.r.t. the single-particle Hamiltonian. ``1d" labels the part where the 
wave functions are distributed linearly over the CPU core. In that 
distribution every thread handles a number of wave functions. In that 
distribution the damped-gradient steps are performed. ``comm" labels 
the time needed to switch between distributions.

The 2d part of the calculation is the most expensive part and is 
expected to scale with $(\text{number of s.p. wave functions})^2$. The 
1d part is expected to be linearly dependent on the number of wave 
functions. In summary, we achieve the expected scaling. Going from the 
system with $L=42\fm$ to the system with $L=58\fm$, the computational 
time increases by a factor of about 7.

\begin{figure}[!t]
    \centering
    \includegraphics[width=0.95\columnwidth]{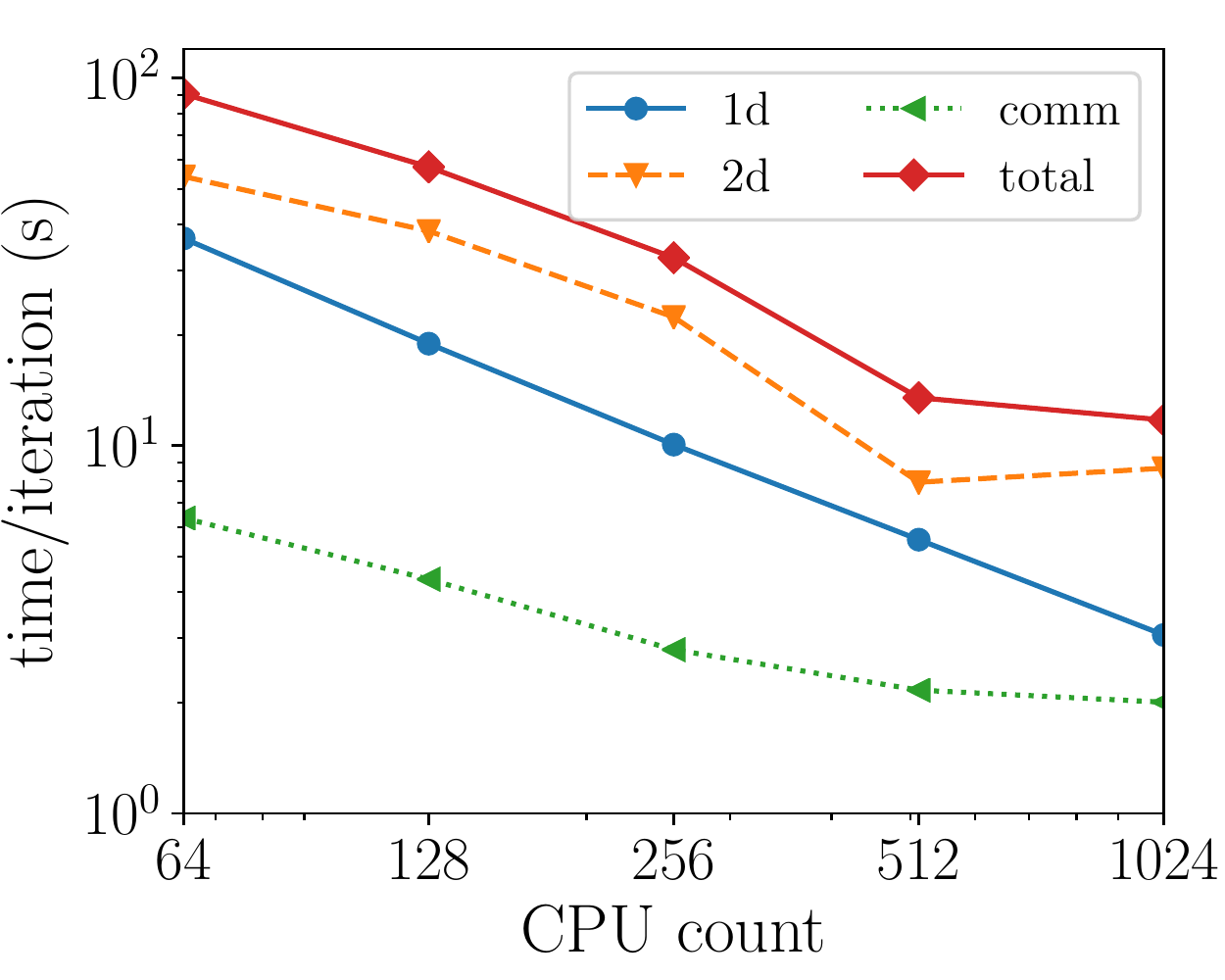}
    \caption{Same as Fig~\ref{fig:scaling_L} but for a system with $L=46\fm$, $\rho=0.08$ and $X_P=1/2$ varying the core count.}
    \label{fig:scaling_cores}
\end{figure}

In Fig.~\ref{fig:scaling_cores}, we show the time per iteration for a fixed system with about 8000 particle wave functions with respect to the core count used. Note that for one node with 32 cores the memory was not sufficient for this large calculation. While the 1d part scales almost perfectly, the dominant 2d part is sensitive to the exact core count. The communication overhead is small and decreases with an increasing number of cores, but becomes comparable to the 1d part for large core counts. Overall, we again obtain a reasonable scaling. In practice, we use mostly 128 cores for smaller cases and 256 or 512 cores for large calculations ($L>40\fm$).


\section{Conclusion}\label{sec:Conlusion}
We studied single and double TPMS pasta configurations in comparison to the slab configuration at intermediate densities with less than half saturation density. As an interaction model, we chose the TOV-min parametrization, which has been fitted also to the mass radius relation of neutron stars. From the large difference of the energy per nucleon of uniform matter to pasta matter at $X_P=1/2$ and $1/3$, we can infer that pasta matter should be realized. Also for $X_P=1/10$ and low densities pasta significantly reduces the energy per nucleon. 

The TPMS configuration cover a large span of different periodic lengths. Although the shapes are very different, physical properties such as the surface area per volume and Euler characteristic per volume reduce to almost the same values in the physical box scales. The energies per nucleon of the different configurations lie within a few hundred keV and thus if the temperature of the system exceeds this difference, several of the configurations can be present at the same time. We therefore expect that pasta matter at finite temperature is not well ordered, but undergoes constant transformation between different configurations and has amorphous character.

The analysis with finite temperature revealed that for a high proton fraction pasta matter can exist with temperatures $T>10\MeV$. For low proton fraction as e.g. present in a neutron star, pasta dissolves at still appreciable but lower temperatures. The preferred box lengths, however, can vary significantly.

We also looked at the impact of electron screening. While the change of the energy per nucleon  is negligible and below the uncertainty of the nuclear interaction model, it can have a small influence on the periodic length of the configurations. Contrary to the common intuition, periodic lengths grow when including electron screening.

\begin{acknowledgments}
  This work was in part supported by the US Department of Energy,
  Office of Science under the award DE-SC0018083 (NUCLEI SciDAC-4
  Collaboration) and the Deutsche Forschungsgemeinschaft (DFG, German
  Research Foundation) - Projektnummer 279384907 - SFB
  1245. Computational resources were provided by the Center for
  Scientific Computing (CSC) of the Goethe University Frankfurt, and
  the National Energy Research Scientific Computing Center (NERSC) in
  Berkeley, California.
\end{acknowledgments}
\bibliography{references}
\end{document}